\documentclass[11pt]{article}

\usepackage[utf8]{inputenc}
\usepackage[T1]{fontenc}
\usepackage[english]{babel}
\usepackage{lmodern}

\usepackage[margin=1in]{geometry}
\usepackage{microtype}

\usepackage{amsmath,amssymb,amsthm}
\usepackage{graphicx}
\usepackage{booktabs}
\usepackage{xcolor}
\usepackage{tabularx}
\usepackage{array}
\usepackage{enumitem}
\usepackage{float}
\usepackage{listings}
\usepackage{cite}
\usepackage{authblk}
\usepackage{hyperref}
\usepackage{cleveref}

\definecolor{codebg}{rgb}{0.97,0.97,0.97}
\definecolor{codekw}{rgb}{0.13,0.29,0.53}
\definecolor{codecm}{rgb}{0.40,0.40,0.40}
\definecolor{codestr}{rgb}{0.64,0.08,0.08}
\hypersetup{
  colorlinks=true,
  linkcolor=black,
  citecolor=black,
  urlcolor=blue!60!black,
  pdftitle={Retrieval-Augmented Generation for Scientific Code Understanding},
  pdfauthor={Aaron Nobile, Andreas Adelmann, and Mohsen Sadr}
}

\theoremstyle{definition}

\theoremstyle{plain}

\title{Retrieval-Augmented Generation for Scientific Code Understanding}
\author[1]{Aaron Nobile\thanks{\texttt{nobilea@ethz.ch}}}
\author[2]{Andreas Adelmann\thanks{\texttt{andreas.adelmann@psi.ch}}}
\author[2,3]{Mohsen Sadr\thanks{\texttt{mohsen.sadr@psi.ch}; \texttt{msadr@mit.edu}}}
\affil[1]{Department of Physics, ETH Zürich, 8093 Zürich, Switzerland}
\affil[2]{Paul Scherrer Institute, 5232 Villigen PSI, Switzerland}
\affil[3]{Massachusetts Institute of Technology, Cambridge, MA 02139, USA}
\date{}

\begin{document}
\maketitle

\begin{abstract}
\noindent
Large language models have become central to modern coding assistants, but
state-of-the-art systems such as Claude Code or Codex~\cite{anthropic2025claudecode,openai2025codex} rely on very large,
cloud-hosted models with significant computational cost and data-privacy
implications. This work investigates whether a useful, fully local
coding agent can be built around small open-source models by shifting the
computational burden away from inference. We develop a Retrieval-Augmented
Generation (RAG) system for scientific code understanding that strictly
separates an expensive offline ingestion stage---parsing, structural graph
construction, LLM-generated entity explanations, and embedding---from a
lightweight online answering stage. The system is evaluated on a 100-question
benchmark spanning eleven categories over the IPPL~\cite{ippl,muralikrishnan2024ippl} scientific codebase written in C++,
with answers scored by an independent frontier model as the judge. Across seven answering
models, we find that model family and retrieval quality matter more than
parameter count, i.e. a 9B model achieves the highest average score (0.795),
outperforming both larger models within our pipeline and the same models
embedded in the Claude Code retrieval architecture. The results indicate that
front-loading code understanding into a reusable, codebase-specialised vector
store enables small local models to deliver grounded and repository-specific
answers, making the agent well suited as a privacy-preserving development tool
for in-house scientific codebases.
\end{abstract}


\section{Introduction}
\label{sec:introduction}
Large language models (LLMs) have become capable assistants for software
development, but the strongest systems generally depend on large,
cloud-hosted models. This creates practical barriers for scientific software:
repositories may be private, inference can be expensive, and the relevant
knowledge is often distributed across years of specialized C++ development,
a challenge also observed in repository-level code generation research~\cite{zhang2023repocoder}.
This work asks whether a smaller, locally executed model can answer technical
questions about such a repository when the surrounding retrieval system does
more of the code-understanding work in advance.

We present a retrieval-augmented system that separates expensive offline
ingestion from lightweight online question answering. The offline stage parses
source and documentation, constructs structural relationships, generates
entity-level explanations, and embeds the resulting records. At query time,
the system combines semantic retrieval, metadata reranking, query-intent
routing, and structural expansion before prompting a local answering model.
We evaluate the approach on 100 questions across eleven categories from the
IPPL scientific C++ codebase. Our contributions are: (i) a code-aware ingestion
pipeline that represents functions, files, modules, and call neighborhoods;
(ii) a retrieval pipeline designed to provide concise, grounded context to
small local models; and (iii) an empirical comparison of seven answering
models and an agentic retrieval baseline.

\section{Background}
\label{sec:background}

\subsection{Large Language Models}
A Large Language Model (LLM) is an advanced artificial intelligence system designed to process and generate sequences of text. These models are trained on massive datasets and are capable of performing a wide range of natural language processing tasks, including text generation, summarization, translation, and question answering. Modern LLMs can also understand and generate source code, making them useful tools for software development~\cite{chen2021codex}.

\begin{enumerate}
\item \textbf{Transformer Architecture}~\cite{vaswani2017attention} - The Transformer is the core architecture behind modern LLMs. It processes text as a sequence of tokens and uses a mechanism called self-attention to determine which tokens are most relevant for understanding the current context. By stacking multiple attention layers, the model can capture complex relationships across long sequences of text.

The Transformer is a major advancement over earlier neural network architectures because it allows all tokens to be processed in parallel, significantly improving computational efficiency and scalability while preserving long-range contextual information.

\item \textbf{Parameters} - Parameters are the trainable coefficients contained within the model. They determine how information flows through the network and how strongly different patterns influence the model's predictions. During training, these parameters learn linguistic structures, semantic relationships, and statistical patterns from the training data. Modern LLMs often contain billions of such parameters.

\item \textbf{Training} - Training is the process through which the model \textit{learns} its parameters. During this process, the model is repeatedly asked to predict missing or subsequent tokens in a text sequence. The prediction error is measured using a loss function, and optimization algorithms such as the gradient descent~\cite{curry1944steepest} adjust the parameters to minimize this error.

Through repeated exposure to large datasets, the model gradually learns patterns in language and develops the ability to generate coherent and contextually relevant responses.

\end{enumerate}

The combination of the Transformer architecture and large-scale training has
enabled language models to address a wide range of language, reasoning, and
programming tasks~\cite{vaswani2017attention,chen2021codex}. This capability has
motivated specialized systems that use LLMs as their central reasoning
component, commonly referred to as AI agents.

\subsection{Coding Agents}
One particularly successful application of LLMs is the coding agent. Rather than acting solely as a chatbot, a coding agent is able to interact with external tools, inspect source code, modify files, execute commands, and reason about software projects. This allows the model to assist developers with tasks such as code generation, debugging, refactoring, and documentation.
\newline
The usefulness of coding agents becomes especially apparent in scientific software projects. Such codebases are often developed over many years, contain complex numerical algorithms, and require substantial domain knowledge in mathematics or physics. Understanding the structure and intent of these projects can therefore be challenging even for experienced developers.
\newline
In recent years, companies such as Anthropic~\cite{anthropic2025claudecode}, OpenAI~\cite{openai2025codex}, Google~\cite{google2025geminicodeassist}, and Cursor~\cite{cursor2025cursor} have developed coding assistants including Claude Code, Codex, and Cursor. These systems rely on very large models hosted on powerful computing infrastructure. While highly capable, they require significant computational resources and are typically associated with usage costs.
\newline
The goal of this work is to investigate whether a useful coding agent can be built around a locally running open-source LLM. Since local models are generally much smaller than state-of-the-art commercial models, the surrounding agent framework must compensate for some of these limitations. By constraining the problem domain and providing structured tools and workflows, the framework aims to enable effective code understanding and generation despite the reduced model size.

\subsection{Retrieval-Augmented Generation (RAG)}

A major limitation of Large Language Models is that their knowledge is limited to the information contained in their training data and the context currently provided in the prompt. In the case of coding agents, this presents a challenge, as software projects often contain thousands of files and many years of development history that cannot be provided to the model at once.
\newline
Retrieval-Augmented Generation (RAG) addresses this problem by combining information retrieval with language generation~\cite{lewis2020rag}. Instead of relying solely on the model's internal knowledge, a RAG system first searches an external knowledge source for information relevant to the user's request and then provides the retrieved information as additional context to the LLM.
\newline
In the context of coding agents, the external knowledge source is typically the source code repository itself. Files, functions, classes, documentation, and comments are indexed and stored in a searchable format. When the user asks a question or requests a code modification, the agent retrieves the most relevant pieces of information from the codebase and includes them in the prompt sent to the model.
\newline
This approach allows a model to work with codebases that are significantly
larger than its context window and to retrieve information from the current
state of the repository rather than relying exclusively on knowledge stored in
its parameters~\cite{lewis2020rag}. These properties motivate our investigation
of whether retrieval can compensate for some limitations of smaller local
models.
\newline
A typical RAG pipeline consists of three main stages. First, documents are processed and converted into vector representations known as embeddings. These embeddings capture the semantic meaning of the code and documentation. Second, a retrieval system performs a similarity search to identify the most relevant documents for a given query. Finally, the retrieved information is appended to the prompt and used by the LLM to generate a response.
\newline
For coding agents, RAG serves as a bridge between the language model and the software project itself. Rather than attempting to memorize an entire codebase, the model can dynamically access the information it needs, improving both accuracy and scalability; repository-level retrieval has likewise been used to expose cross-file context to code models~\cite{zhang2023repocoder}. This principle forms the basis of the retrieval system presented here.

\subsection{IPPL as a Scientific-Software Testbed}

The Independent Parallel Particle Layer (IPPL) is an open-source C++ framework
for developing performance-portable Eulerian, Lagrangian, and hybrid
particle-mesh applications~\cite{ippl,muralikrishnan2024ippl}. Its use of MPI,
Kokkos for performance-portable on-node parallelism~\cite{edwards2014kokkos},
and heFFTe for distributed Fourier transforms~\cite{ayala2020heffte} makes it a
representative testbed for repository-level questions that combine software
structure with high-performance-computing and numerical concepts.


\section{System Design}
\label{sec:method}
The system is a locally executable scientific-code assistant built from
open-source language-model components. Because repository-specific knowledge
resides primarily in the codebase itself, the design emphasizes retrieval that
supplies the answering model with structured, precise, and relevant context.

\subsection{System Overview}
\begin{figure}[H]
    \centering
    \includegraphics[width=\linewidth]{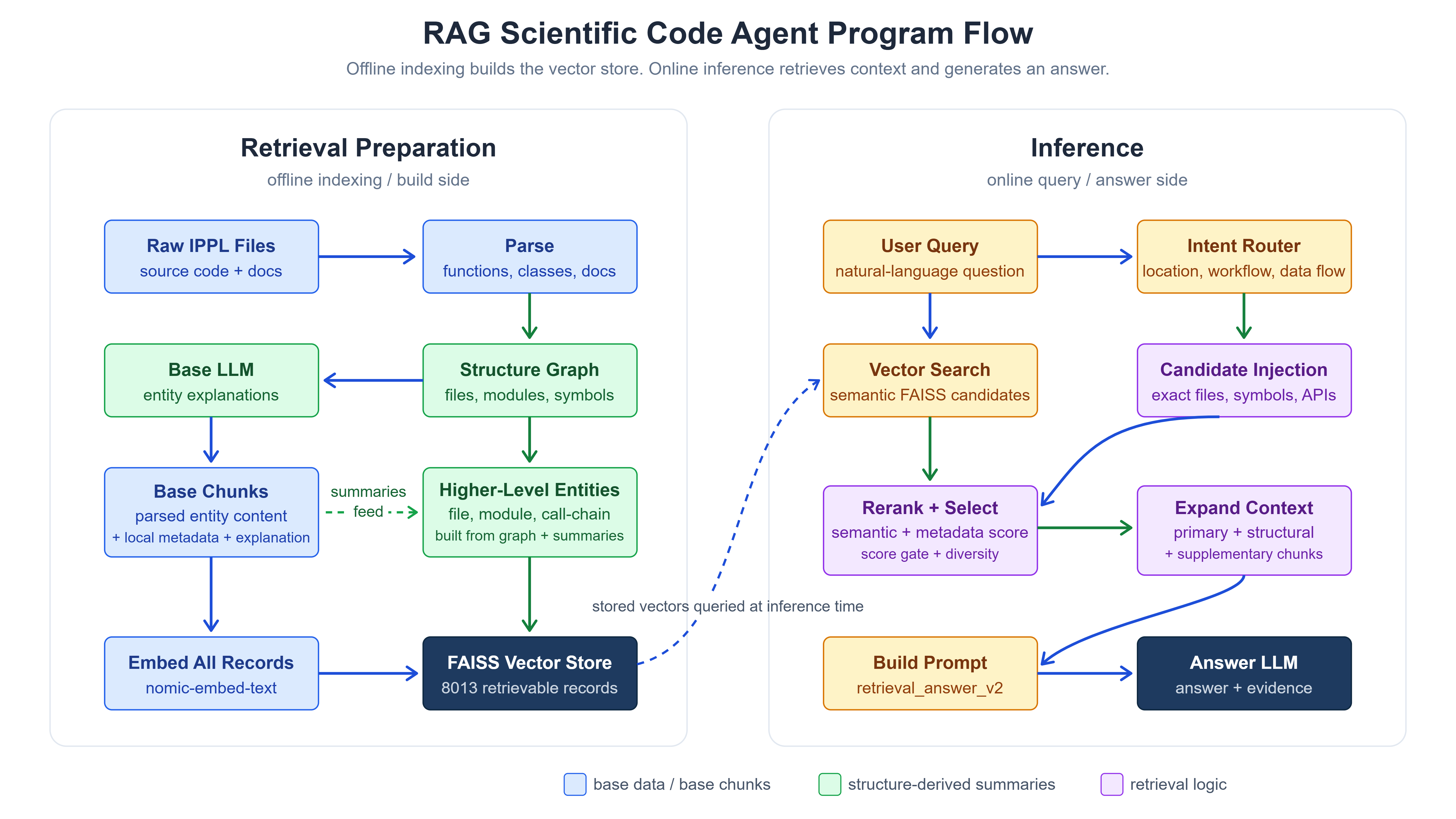}
    \caption{The flowchart of the RAG-Scientific-Code-Agent can be split into two parts: Retrieval Preparation in which we setup Structure Graphs, Base LLM explanations and Higher-Level Entities to save them as retrievable chunks in a vector store, and Inference which focuses on retrieval in a direct sense, vector search, query intent detection, candidate injection, reranking, context expansion and prompt building.}
    \label{fig:system-overview}
\end{figure}

The architecture uses \textit{prompt enrichment}: structured excerpts from the
codebase, called \textit{chunks}, are inserted into a prompt template before it
is sent to the LLM. This design cleanly separates retrieval preparation from
inference. \Cref{fig:system-overview} summarizes both stages.

\subsection{Ingestion and Retrieval Preparation (Offline)}

The first stage of the pipeline consists of ingesting all relevant files from the target codebase. During this offline preprocessing step, the developed file extraction module \texttt{file\_reader.py} traverses the project directory and collects source code and documentation files that may contain information relevant to subsequent retrieval tasks. The following file types are considered:

\begin{enumerate}[itemsep=0pt]
    \item Source code files: \texttt{.cpp}, \texttt{.hpp}, and \texttt{.h}
    \item Documentation files: \texttt{.md}, \texttt{.rst}, and \texttt{.txt}
\end{enumerate}
\ \newline
The extracted files serve as the main building block from which relevant chunks are formed.

\subsubsection{Parsing and Chunking}
Since there are multiple different file types for C++ code (\texttt{.cpp}, \texttt{.hpp}, and \texttt{.h}) and documentation (\texttt{.md}, \texttt{.rst}, and \texttt{.txt}) the pipeline requires different parsers.
\newline
The main parser for \texttt{.cpp} files is \texttt{cpp\_parser.py}, which parses the implementation files of the C++ Codebase. Therefore this file takes in raw \texttt{.cpp} files from the source code and determines what the system \textit{sees} at the source level function entity. Each extracted function record contains metadata like:
\begin{lstlisting}[language=Python,caption={Extracted metadata from function parsing},label=lst:function-metadata]
{
    "path": ...,
    "file_name": ...,
    "base_name": ...,
    "source_type": ...,
    "entity_type": ...,
    "chunk_type": ...,
    "symbol_name": ...,
    "parent_symbol": ...,
    "class_name": ...,
    "return_type": ...,
    "parameters": ...,
    "leading_comment": ...,
    "code": ...
}
\end{lstlisting}
This metadata matters a lot for retrieval (\cref{sec:retrieval}) later on, because reranking (\cref{sec:metadata-reranker}) uses fields like 'symbol\textunderscore name', 'path', 'source\textunderscore type', 'chunk\textunderscore type', and 'parent\textunderscore symbol'. 
\newline
The core tool used to analyze C++ syntax is \texttt{TreeSitterParser}, built on
Tree-sitter's incremental concrete-syntax-tree parser~\cite{treesitter2026} and
the C++ grammar distributed by \texttt{tree\_sitter\_languages}. Unlike regular
expression matching, this approach represents C++ syntax structurally.

\begin{lstlisting}[language=Python,caption={Example of a detected function from \texttt{cpp\_parser.py}},label=lst:parsed-function]
void ippl::initialize(int& argc, char* argv[], MPI_Comm comm) {
    ...
}
\end{lstlisting}

This returns: 
\begin{lstlisting}[language=Python,caption={Example metadata extracted during function parsing},label=lst:parsed-function-metadata]
symbol_name: initialize
parent_symbol: ippl
return_type: void
parameters: int& argc, char* argv[], MPI_Comm comm
code: full function body
\end{lstlisting}
\ \newline
The developed file \texttt{cpp\_parser.py} contains several functions responsible for processing and organizing parsed source code entities. One such function, \texttt{\_split\_qualified\_name(...)}, separates fully qualified names by removing the C++ scope resolution operator. For example, the function signature \texttt{void ippl::initialize()} is split into the namespace \texttt{ippl} and the function name \texttt{initialize}. This separation simplifies subsequent indexing and retrieval tasks.
\newline
Another important component of the parser is the \texttt{CommentExtractor} function. This utility associates leading comments with the corresponding parsed entities, such as functions or classes. By preserving developer-provided documentation, the parser retains contextual information that would otherwise be unavailable during retrieval.
\\
The developed \texttt{header\_parser.py} module is responsible for parsing C++ header files (\texttt{.h} and \texttt{.hpp}). Its primary purpose is to extract declarations and structural information, including function signatures, class definitions, namespaces, and other publicly exposed interfaces. In contrast, \texttt{cpp\_parser.py} processes implementation files (\texttt{.cpp}) and extracts the corresponding function and method implementations. Consequently, the header parser captures the interface of the codebase, whereas the C++ parser captures its implementation.

\sloppy The primary role of the header parser is to transform the contents of header files into structured representations of the codebase. The extracted entities include
    Namespaces, Classes, Structs, Method declarations \& definitions and Qualified free-function declarations \& definitions.

The Header Parsing is implemented in the \texttt{TreeSitterHeaderParser} class, it uses \texttt{TreeSitterParser} and a CommentExtractor just like in the cpp parser. Generally the Header Parser looks for:
\begin{enumerate}
    \item type\textunderscore identifier / identifier
    \item field\textunderscore declaration\textunderscore list
    \item base\textunderscore class\textunderscore clause
\end{enumerate}
Then returns a record like:
\begin{lstlisting}[language=Python,caption={Record produced by the header parser},label=lst:header-record]
{
    "chunk_type": "class" or "struct",
    "symbol_name": class_or_struct_name,
    "parent_symbol": class_or_struct_name,
    "return_type": "class" or "struct",
    "parameters": inheritance_clause,
    "code": full class/struct source,
}
\end{lstlisting}

\subsubsection{Documentation chunking \& parsing}
For documentation parsing, all documentation files within the codebase are processed by \texttt{doc\_parser.py}. The parser identifies and extracts logical sections from Markdown, reStructuredText, and plain text documents.
\newline
These sections do not yet correspond to the final entries stored in the vector database. Instead, they are passed to the \texttt{DocChunker}, which converts them into structured chunk records containing metadata such as \texttt{symbol\_name}, \texttt{chunk\_type}, \texttt{code\_block}, and \texttt{entity\_level}. To ensure that individual chunks remain within a predefined size limit, the \texttt{DocChunker} further subdivides sections whose content exceeds the configured \texttt{max\_chunk\_size}.
\newline
Each resulting chunk is represented as a retrievable record of the form:

\begin{lstlisting}[language=Python,caption={Documentation chunk record},label=lst:documentation-chunk]
{
"source_type": "documentation",
"symbol_name": section_title,
"chunk_type": section_type,
"language": "markdown" / "rst" / "text",
"section_path": ...,
"code": chunk_content,
"entity_type": section_type,
"entity_level": "documentation_section_level",
}
\end{lstlisting}
Since the chunker enforces a maximum chunk size, a single documentation section may be divided into multiple chunks. In such cases, the corresponding section title is extended with a numerical suffix (e.g., \texttt{section\_title\_1}, \texttt{section\_title\_2}, \ldots) to preserve the relationship between the generated chunks and their original section.

\subsubsection{Call and Dependendency Graph Building}
Large Code bases are often time concatenated with multiple functions, files and folders. In addition, large scientific simulations also contain various call-chains. The Language model needs to retrieve these correctly in order to deliver a good explanation. \texttt{project\_structure\_builder.py} inside \texttt{src/structure} tries to solve this issue by building several "structure graphs" after parsing. \texttt{ProjectStructureBuilder} combines them into one \texttt{project\_structure}. The main output is stored inside \texttt{embeddings/project\_structure/project\_structure.json}.

\subsubsection{Higher Entity Building}
As of now retrieval preparation only creates chunks based on code structures hence one chunk is a entire function, a class, a namespace etc... This function-level chunking technique is very common and ensures that upon retrieval there is not much noise inside the retrieval context. However what happens if the user query asks about an entire Module? 

\begin{lstlisting}[caption={Example query for which a summary of the FFT module should be retrieved},label=lst:module-query]
    What files are inside the FFT module in the source code? 
\end{lstlisting}

A query like this faces a problem, even if all the functions inside the files of the FFT module would get retrieved, which is hardly possible, the context-size would flood the LLM with noise and deliver a wrong result. A solution however to this problem is building a specific "Higher Entity Chunk" specifically for the module level in this case the FFT module. This is exactly what the program does. Using the project structure alongside the file-contents of the FFT module the program creates a structured list of contents which get sent to a LLM to create a specfic Module level chunk. It is reasonable that this steps also occurs also on the file-level stage.
\newline
In \texttt{runtime\_settings.json}, the user can choose different entity building strategies. There are three settings, which can be adjusted:

\begin{table}[H]
\centering
\begin{tabularx}{\linewidth}{l X}
\hline
\textbf{Setting} & \textbf{Meaning} \\
\hline
\texttt{file\_level\_entity\_strategy} & How file-level summary entities are created. \\
\texttt{module\_level\_entity\_strategy} & How module/folder-level summary entities are created. \\
\texttt{call\_chain\_entity\_strategy} & How call-chain/workflow entities are created. \\
\hline
\end{tabularx}
\caption{Entity-building strategy settings, enlisted in the runtime-settings.}
\label{tab:strategy-settings-meaning}
\end{table} 
\texttt{unified\_whole\_file\_with\_fallback\_v1} creates one file-level entity per file. It uses parsed symbols and their explanations when available. If no symbols were parsed, it falls back to visible whole-file/raw content evidence.
\newline
\texttt{module\_entities\_from\_descendant\_files\_v1} creates module-level entities from the files inside a module/folder. It reads lower file summaries and structural relationships to describe what the module is responsible for.

\texttt{direct\_call\_neighborhood\_v1} creates call-chain entities around callable symbols. It uses direct incoming and outgoing call relationships, plus symbol/file/module summaries, to explain a local workflow neighborhood.

Therefore the project structure has three main node-like lists: "symbols" (functions, classes, structs,...), "files" and "modules" (folders and subfolders)

\subsubsection{Embedding and Vector Store}

The embedder, implemented in \texttt{embedder.py}, serves as the interface between the FAISS vector store~\cite{johnson2019faiss} and the retrievable content chunks. It has two primary responsibilities:

\begin{enumerate}
\item Embedding the chunked content.
\item Embedding the user query.
\end{enumerate}

The overall purpose of the embedder is to transform both the stored content and the user's question into vector representations such that semantically related content yields a high similarity score during retrieval. When content chunks are embedded, the pipeline incorporates their parsed information and associated metadata into a structured representation. This process is implemented in \texttt{\_build\_chunk\_embedding\_prompt} and follows a predefined format.

\begin{lstlisting}[language=Python,caption={Content chunk structure generated inside \texttt{embedder.py}},label=lst:embedding-chunk]
File: ...
Symbol: ...
Chunk Type: ...
Parent Symbol: ...
Section Path: ...
Generated Explanation:
...
Code:
...
\end{lstlisting}

This design is a distinguishing feature of the pipeline, as both semantic explanations and structural metadata become part of the final vector representation. In addition, the embedder applies a \texttt{prompt\_limit} to the generated embedding prompt. During fallback retries, the fields \texttt{generated\_explanation} and \texttt{code} may be truncated to satisfy the context length constraints of the embedding model. Consequently, if an embedding attempt fails due to excessive prompt length, a second attempt is performed using a shortened version of the chunk.

Since the embedder is also responsible for embedding user queries, it contains initial mechanisms for query intent detection. Before vectorizing a user query, the embedder searches for file name patterns using a regular expression.

\begin{lstlisting}[language=Python,caption={Regular expression used to detect file names such as \texttt{Ippl.cpp} or \texttt{BareField.h} in a user's query},label=lst:file-pattern]
self.file_extension_pattern = re.compile(
r"\b[A-Za-z0-9_-]+.(?:cpp|hpp|h|md|rst|txt)\b",
re.IGNORECASE,
)
\end{lstlisting}

If a file name is detected in the user query, the embedder explicitly incorporates this information into the query embedding prompt generated by \texttt{\_build\_query\_embedding\_prompt}. The objective is to structure the query in a similar manner to the embedded content chunks, thereby increasing the likelihood that relevant chunks are retrieved.

It is important to note that this prompt structure is only used for generating the query embedding. The original user query sent to the LLM remains unchanged.

\begin{lstlisting}[language=Python,caption={Query embedding prompt template implemented in \texttt{embedder.py}},label=lst:query-embedding-prompt]
def _build_query_embedding_prompt(self, text):
file_names = self._extract_file_names(text)
file_name = ", ".join(file_names)
intent = (
"find relation across FILES which is asked in QUESTION"
if len(file_names) > 1
else self._infer_intent(text)
)
chunk_types = self._find_chunk_type(text)
chunk_type = ", ".join(chunk_types) if chunk_types else "any"
detected_code = self._extract_detected_code(text)

```
    sections = []
    if file_name:
        sections.append(f"File: {file_name}")
    if chunk_type:
        sections.append(f"Chunk Type: {chunk_type}")
    if intent:
        sections.append(f"Intent: {intent}")
    sections.append(f"Question:\n{text}")
    if detected_code:
        sections.append(f"Code:\n{detected_code}")
```

\end{lstlisting}

Depending on the content of the user query, information such as file names, chunk types, inferred intent, and detected code fragments is extracted and incorporated into the embedding prompt. This allows the query embedding to more closely resemble the structure of the indexed content chunks, improving the effectiveness of dense retrieval before reranking is applied.

\subsubsection{Vector Store}
\label{sec:vector-store}
The vector store is the "library" in which all vectorized content chunks get stored and where \texttt{retriever.py} fetches all Top-k chunks from it. The vector store gets built by the python script \texttt{vector\_store.py}. Since generating embeddings and building the vector store is computationally demanding the vector store relies on a persisted manifest which signals the pipeline to not rebuild the vector store if unchanged.

\begin{lstlisting}[language=Python,caption={Persisted vector-store manifest used to identify reusable artifacts},label=lst:vector-manifest]
embedding_backend: ollama
embedding_model: nomic-embed-text
dimension: 768
metadata_count: 8013
vector_count: 8013
max_chunk_size: 400
\end{lstlisting}

For both searching and storing vectors the pipeline uses FAISS~\cite{johnson2019faiss}. When the \texttt{VectorStore} class is initialized, its FAISS index is configured for exact L2-distance search. All embeddings are converted to 32-bit floating-point values before insertion.

Upon entering a specific user question the \texttt{search} function inside \texttt{vector\_store.py} uses the embedded query vector to find top-k closest chunks. Each top-k retrieved chunk will be stored inside results array alongside the chunk content and the determined distance. 

\begin{lstlisting}[language=Python,caption={Search in \texttt{vector\_store.py} returns the top-$k$ chunks and their distances from the query vector},label=lst:vector-search]
    def search(self, query\_vector, k=5):
        #embeded prompt of user = query\_vector -> turn to float32
        #k...amount of neigherst search results returned.
        query_vector = np.array([query\_vector]).astype("float32")
        #distances = vector L2 distances away from the vector (rising)

        distances, indices = self.index.search(query\_vector, k)
        results = []
        for distance, index in zip(distances[0], indices[0]):
            if index < 0 or index >= len(self.metadata):
                continue
            results.append(
                {
                    "chunk": self.metadata[index],
                    "distance": float(distance),
                }
            )
\end{lstlisting}

All previously shown implementations follow Attention based dense retrieval search only, however hybrid retrieval is important since dense retrieval can miss direct keyword search. For this the vector store appends the retrieved chunk list by using three functions: \texttt{get\_chunks\_by\_filenames}, \texttt{get\_chunks\_by\_symbols} and \texttt{get\_chunks\_by\_modulenames}.

The vector store provides several exact-retrieval helpers in addition to normal semantic vector search. \texttt{get\_chunks\_by\_filenames} retrieves file-level entities for explicitly mentioned files. It distinguishes between a full file name, such as \texttt{BareField.hpp}, and a base name without extension, such as \texttt{BareField}. If only the base name is mentioned, all existing file-level entities with that base name are injected, e.g. \texttt{BareField.h} and \texttt{BareField.hpp}, with distance \texttt{0.0}. If a full file name is mentioned, the exact file receives distance \texttt{0.0}, while sibling files with the same base name but different extensions are also added using their normal vector distance.

\texttt{get\_chunks\_by\_symbols} retrieves function-level chunks whose symbol name exactly matches symbols detected in the query. To improve precision for queries that mention both a function and its containing file or class, the method checks whether the matched chunk’s \texttt{base\_name}, file stem, or full \texttt{file\_name} also appears in the query-derived symbols. Such contextual matches receive distance \texttt{0.0}, while broader file-subject matches are kept slightly lower priority. Only the top two symbol candidates are returned.

\texttt{get\_chunks\_by\_modulenames} retrieves module-level entities by exact module name. It is intended to provide a direct retrieval path for questions that explicitly mention a module or subsystem, returning matching module-level chunks with distance \texttt{0.0}.

\texttt{search\_in\_filenames} performs a filename-scoped semantic search. Unlike exact file injection, it computes vector distance between the query and chunks inside the given files. It is restricted to \texttt{function\_level} chunks and returns the top three closest matches, which helps retrieve the most relevant functions or methods from already identified files.

The whole vector store in the end gets saved under \texttt{embeddings/vector\_store} therefore without changing the manifest the pipeline calls up the already stored \texttt{vector\_store} when retrieving chunks.

\subsection{Retrieval}
\label{sec:retrieval}
Retrieval is the component of the system responsible for selecting the most relevant pieces of information from the indexed codebase before an answer is generated. Instead of sending the entire project to the language model, the retrieval pipeline searches over embedded code chunks, structural metadata, symbol names, file paths, and higher-level entity descriptions to identify a compact context that is likely to contain the answer. This step is essential because large codebases exceed the context window of an LLM and contain many unrelated files. By ranking and filtering candidate chunks according to semantic similarity, lexical matches, and entity-level relevance, retrieval acts as the bridge between the user’s question and the parts of the repository that the model should reason about. A strong retrieval stage therefore directly improves answer quality: it reduces noise, surfaces the right implementation details, and makes the final response more grounded in the actual source code.

\subsubsection{Chunk Classification}
After the candidate chunks have been retrieved, they are categorized into three groups based on their role within the retrieval pipeline:

\begin{enumerate}
    \item \textbf{\texttt{primary}:} main chunks selected directly from the candidate pool after semantic search, exact filename/symbol injection, re-ranking, score gating and diversification. They are the core answer evidence

    \item \textbf{\texttt{structural\_expansion}:} chunks added after primary retrieval by \texttt{structural\_expander.py}. They are structurally related to the primary chunks e.g. file-level chunk can expand module-level chunk; function-level chunk can expand call-chain chunk etc.

    \item \textbf{\texttt{supplementary}:} chunks retrieved from files referenced by the primary chunks. The retriever collects \texttt{referenced\_files} from \textbf{primary chunks}, builds a supplementary query and searches inside those referenced files using \texttt{search\_in\_filenames(...)}
\end{enumerate}

\subsubsection{Retriever}
The central component that orchestrates retrieval is \texttt{retriever.py}. Its primary task is to decide which retrieved content chunks are ultimately passed to the LLM and which candidates are filtered out. This functionality overlaps conceptually with parts of the vector store described in \autoref{sec:vector-store}, but the two components operate at different stages of the retrieval process. The functions in \texttt{vector\_store.py} are responsible for storing chunks and computing their initial relevance to a query, including adjustments to the retrieval distance for exact file-name or symbol-name matches. In contrast, \texttt{retriever.py} works with the ranked candidates returned by the vector store and decides how these candidates should be combined, filtered, and balanced before being sent to the LLM.

In this sense, the vector store answers the question: ``Which chunks exist, and how close are they to the query?'' The retriever answers the follow-up question: ``Which mixture of chunks should the LLM actually see?'' This makes the retriever the final selection layer between similarity search and answer generation.

For choosing the final candidates, the retriever uses several configurable parameters during the initialization of the \texttt{Retriever} class. In particular, it defines five adjustable settings that control the selection of primary chunks.

\begin{table}[H]
    \centering
    \small
    \begin{tabularx}{\linewidth}{>{\raggedright\arraybackslash}p{0.32\linewidth} c X}
        Setting & Value & Description\\
        \hline
        \texttt{PRIMARY\_SCORE\_RELATIVE\_FLOOR} & 0.4 & Controls how far below the best reranked candidate may fall before it is considered weak. Used in \texttt{\_select\_primary\_candidates(...)}\\
        \texttt{PRIMARY\_SCORE\_GAP\_THRESHOLD} & 8.0 & Controls how large a score drop between two consecutive accepted candidates must be before the retriever treats it as a quality cliff. The value states the difference; as soon as "diff" $\geq$ \texttt{PRIMARY\_SCORE\_GAP\_THRESHOLD} the selection stops if the candidate is also below the relative floor.\\
        \texttt{PRIMARY\_SCORE\_ABSOLUTE\_FLOOR} & 1.25 & Defines minimum absolute score threshold. Below this threshold $\rightarrow$ very weak candidate.\\
        \texttt{PRIMARY\_SAME\_SYMBOL\_CAP} & 1 & Limits how many primary chunks with same symbol group (file path + symbol name) we can retrieve.\\
        \texttt{PRIMARY\_SAME\_FILE\_CAP} & 2 & Limits how many non-file-level primary chunks with the same filename we can retrieve.\\
        \texttt{DOMINANT\_FILE\_RATIO} & 0.75 & Detects when the selected primary candidates are mostly from one file, the value describes a threshold value for \texttt{dominant\_count / total\_count}. Meaning: exceeding the threshold has to many chunks from one file only.
    \end{tabularx}
    \caption{Constants to control which final primary chunks the retriever chooses. They are mainly used to avoid weak or repetitive context and can be adjusted manually in the config \texttt{runtime\_settings.json}.}
    \label{tab:retriever-settings}
\end{table}

These Parameters get used by the main function of the retrieval pipeline: \texttt{retrieve\_with\_diagnostics(self, query, k=5) ...} . The function takes the user query and returns:
\begin{lstlisting}
    {
    "chunks": final_retrieved_chunks,
    "diagnostics": debug_information
}
\end{lstlisting}

So it does both retrieve context for the answer LLM and record why those chunks were selected. In addition the function extracts query signals such as: 

\begin{lstlisting}
    exact_filenames
    exact_symbols
    api_bearing_terms
    data_flow_direction
    data_flow_terms
    comparison_subjects
\end{lstlisting}

Next it calls \texttt{QueryIntentRouter} to classify the user query, into one of the following categories:

\begin{lstlisting}
    api_usage
    location_lookup
    data_flow
    comparison
    symbol_explanation
    file_purpose
    module_overview
    workflow_explanation
    default
\end{lstlisting}

After extracting both the categoric terms with the reranker and the query intent with the \texttt{QueryIntentRouter}, the function calls up the implemented FAISS vector search implemented in \texttt{vector\_store.py}. The \texttt{candidate\_count} is determined both by the selected candidate number \texttt{k} and the \texttt{self.candidate\_k=20} set, which can be adjusted in the config. The semantic candidates are chosen from the search function inside the vector store file. Then exact filename/symbol targets and candidates are acquired by the corresponding functions inside the vector\_store. For comparison intent the retriever uses a specific \texttt{self.\_retrieve\_comparison\_candidates(...)} function to prevent flooding the candidate pool with only one side of mentioned exact key-name hits.

\begin{table}[H]
\centering
\small
\begin{tabularx}{\textwidth}{lXl}
\hline
\textbf{Candidate source} & \textbf{Purpose} & \textbf{Active when} \\
\hline
Semantic candidates &
Top-$k$ vector-search results from the full vector store. This is the broad semantic retrieval baseline. &
Always \\

Exact filename candidates &
Injects file-level chunks for explicitly mentioned filenames or basename matches, e.g. \texttt{BareField} matching \texttt{BareField.h} and \texttt{BareField.hpp}. &
Always \\

Exact symbol candidates &
Injects function-level chunks whose symbol name exactly matches query symbols, e.g. \texttt{getAllocated}. &
Non-comparison queries \\

Subject file candidates &
Injects file-level chunks whose filename stem or basename matches a query subject. This anchors queries around files/classes mentioned without extension. &
Non-comparison queries \\

Comparison candidates &
Injects balanced anchors for each compared subject, so both entities in a comparison receive evidence. &
Comparison queries \\

Literal API candidates &
Scans chunk text for concrete API calls such as initialization/finalization calls and injects exact call-site evidence. &
API/location queries \\

Target-aligned candidates &
Injects chunks matching an explicitly requested entity level, such as file-level, module-level, function-level, or call-chain-level. &
When entity target is detected \\
\hline
\end{tabularx}
\caption{Candidate pool sources used by the retrieval pipeline before reranking and primary chunk selection.}
\label{tab:candidate-pool-overview}
\end{table}

After enlisting all possible candidates seperately \texttt{\_merge\_candidates(...)} concatenates all candidate sources. It priotizes injected candidates by order, semantic vector candidates last. Then removes all duplicate chunks. The actual scoring/ranking happens afterward in the reranker. After re-ranking the pipeline the candidates get refined a final time. \texttt{\_select\_primary\_candidates} removes weak reranked primary candidates. \texttt{\_refine\_primary\_candidates} applies query-intent logic and diversify and \texttt{\_ensure\_exact\_filename\_chunks} implements final exact-match ordering safeguard.

\subsubsection{Test/Build Noise Sorting}
Since the vector store contains build and test files in addition to implementation files from the source code, the program needs to distinguish between these categories during retrieval. Test and build files are usually not part of the core implementation; instead, they often demonstrate, validate, or configure functionality that appears elsewhere in the codebase. The constant \texttt{TEST\_MODULE\_SCOPES} defines which module scopes the retriever treats as test or build noise, unless the user explicitly asks about tests or build-related files.

\begin{lstlisting}
    TEST\_MODULE\_SCOPES = frozenset({"test", "tests", "unit_test", "unit_tests"})
    BUILD_FILE_NAMES = frozenset({"cmakelists.txt"})
    TEST_QUERY_KEYWORDS = (
    "test",
    "tests",
    "unit test",
    "unit_test",
    "unittest",
    "cmake",
    "cmakelists",
    "build",
    "integration test",
)
\end{lstlisting}

This constant lists the module scopes that are considered test- or build-related. A chunk is treated as belonging to this category if its metadata contains a module scope such as:
\begin{lstlisting}
module_scope = "test, tests, ..."
\end{lstlisting}

or:

\begin{lstlisting}
module_scope = "unit_tests, cmake, cmakelists, ..."
\end{lstlisting}

The corresponding filtering logic can be summarized as follows:
\begin{lstlisting}
The retrieved chunk comes from a test or build file
AND the query does not mention tests or build configuration
THEN remove that chunk from the final context.
\end{lstlisting}

However, if the user explicitly asks a test- or build-related question, for example:
\begin{lstlisting}
Where is this tested?
How is the build configured?
What does CMakeLists.txt do?
\end{lstlisting}
then chunks whose module scope is listed in \texttt{TEST\_MODULE\_SCOPES} remain eligible for retrieval. This prevents test and build files from polluting the context of general implementation-focused answers, while still allowing them to be retrieved when they are directly relevant to the query.

\subsubsection{Structural Expansion}
Structural expansion is a method used in the pipeline of the retriever by calling 
\begin{lstlisting}
    structural_result = self.structural_expander.expand(
    query,
    primary_chunks,
    mode=intent_result["structural_mode"],
)
\end{lstlisting}

Structural expansion is applied after the primary retrieval step. First, the retriever collects candidates, reranks them, selects the primary chunks, and applies exact-match safeguards. Only then does it call the structural expander on the selected primary chunks. The expansion mode is chosen by the query intent, for example API-usage queries use a narrower expansion profile, while data-flow and workflow queries use a broader profile. The expander does not perform a new semantic search; instead, it follows structural metadata relationships that were built during ingestion. Depending on the entity level of a primary chunk, it can add related file-level, module-level, call-chain-level, or symbol-level chunks. For example, a selected function-level chunk can be expanded to its file-level summary or module-level summary if the active profile allows it. Each expansion profile limits which relationships are followed and how many additional chunks may be added. Thus, structural expansion enriches trusted primary evidence with nearby structural context while keeping the amount of added context bounded.

The number of added chunks is capped by the active profile:
\begin{lstlisting}
    mode_api_usage   -> max 3 expansions
    mode_1_minimal   -> max 2 expansions
    mode_2_balanced  -> max 3 expansions
    mode_3_broad     -> max 4 expansions
\end{lstlisting}

The conditions to structurally expand are:
\begin{lstlisting}
    1. There must be primary_chunks.
    2. The QueryIntentRouter must assign a structural_mode.
    3. The StructuralExpander must find related chunks in its indexes.
    4. The expansion profile must allow that kind of relationship.
    5. The related chunk must not already be one of the primary chunks.
    6. The max_total_expansions limit must not be reached.
\end{lstlisting}

\subsubsection{Query Intent Routing}
\label{sec:query-intent-routing}

Query intent routing describes the process of identifying the purpose of a user query before the final retrieval context is assembled. This is an important part of the retrieval pipeline because the amount and type of structural expansion should depend on what the user is asking for. For example, a location lookup should retrieve a narrower context than a workflow or data-flow question, which usually requires broader surrounding information.

The current implementation in \texttt{query\_intent\_router.py} categorizes queries into the following intent classes:

\begin{table}[h]
\centering
\begin{tabular}{ll}
\hline
\textbf{Query intent} & \textbf{Structural expansion mode} \\
\hline
\texttt{api\_usage} & \texttt{mode\_api\_usage} \\
\texttt{location\_lookup} & \texttt{mode\_1\_minimal} \\
\texttt{data\_flow} & \texttt{mode\_3\_broad} \\
\texttt{comparison} & \texttt{mode\_2\_balanced} \\
\texttt{symbol\_explanation} & \texttt{mode\_2\_balanced} \\
\texttt{file\_purpose} & \texttt{mode\_2\_balanced} \\
\texttt{module\_overview} & \texttt{mode\_2\_balanced} \\
\texttt{workflow\_explanation} & \texttt{mode\_3\_broad} \\
\texttt{default} & \texttt{mode\_2\_balanced} \\
\hline
\end{tabular}
\caption{Query intent categories and their corresponding structural expansion modes.}
\label{tab:query-intent-categories}
\end{table}

In addition to assigning an intent, \texttt{query\_intent\_router.py} also detects whether the query explicitly targets a certain entity level. Keywords such as ``file'', ``module'', ``call chain'', ``function'', and ``documentation section'' are translated into concrete retrieval preferences. These preferences tell the retriever which entity levels or chunk types should be preferred when ranking the final context.

The constant \texttt{SYMBOL\_LEVEL\_CHUNK\_HINTS} follows the same principle for symbol-level queries. It maps query terms such as ``class'', ``struct'', ``method'', ``function'', ``routine'', and ``namespace'' to more specific chunk types. For example, ``function'' is mapped to \texttt{function\_definition} and \texttt{method\_definition}, while ``namespace'' is mapped to \texttt{namespace}. These hints are only applied when the query targets symbol-level information.

The \texttt{route(...)} function is responsible for deciding which kind of question the user asked and which structural expansion mode and retrieval preferences should be used. The function proceeds as follows:

\begin{enumerate}
    \item Detect whether the query explicitly asks for a specific entity type using \texttt{\_detect\_entity\_target(...)}.
    \item Detect the query intent, for example API usage, location lookup, data flow, comparison, workflow explanation, or module overview.
    \item Return the detected intent, structural expansion mode, routing reasons, entity target, and retrieval preferences.
\end{enumerate}

\subsubsection{Metadata Reranker}
\label{sec:metadata-reranker}

The file \texttt{reranker.py} implements the metadata reranker used inside the retrieval pipeline. It takes the candidate chunks returned by the vector store and reranks them before the structural expansion step. As described in \autoref{sec:vector-store}, the vector store returns an L2 distance between the query embedding and each retrieved chunk embedding. The reranker converts this distance into a semantic score and combines it with metadata-based signals that describe how well a chunk matches the query on the level of file names, symbols, namespaces, source types, API calls, data-flow terms, and requested entity types.

\begin{lstlisting}[language=Python,caption={Scoring formula for the reranked candidate score. The parameters \texttt{lexical\_metadata\_weight} and \texttt{entity\_target\_weight} are configurable in \texttt{runtime\_settings.json}.},label=lst:metadata-reranking-score]
semantic_score = 1.0 / (1.0 + max(distance, 0.0))

metadata_score = (
    lexical_metadata_weight * lexical_metadata_score
    + entity_target_weight * entity_target_score
)

combined_score = semantic_score + metadata_score
\end{lstlisting}

The \texttt{lexical\_metadata\_score} captures direct textual and structural overlap between the query and a candidate chunk. It rewards exact filename matches, exact symbol matches, namespaced symbol matches, namespace matches, and overlap with file paths or symbol names. It also considers whether the query explicitly asks for a header, source, or documentation file and whether the chunk has the corresponding \texttt{source\_type}.

The \texttt{entity\_target\_score} captures whether the chunk matches the type of entity requested by the query intent router. For example, if the user asks for a module, module-level chunks are boosted; if the user asks for a file, file-level chunks are preferred. This prevents lexically related but structurally wrong chunks from dominating the final ranking.

The reranker also contains several query-specific boosts. For API or location queries, it extracts API-bearing terms such as \texttt{Kokkos::initialize} and rewards chunks that contain the corresponding call sites. For data-flow questions, it detects directions such as grid-to-particles or particles-to-grid and boosts chunks containing terms such as \texttt{gather}, \texttt{scatter}, \texttt{interpolation}, or related particle/grid terminology. For comparison questions, it extracts the compared subjects and rewards chunks whose metadata or text matches one of those subjects.

In addition, \texttt{reranker.py} includes safeguards against misleading matches. Very low-signal symbol names such as \texttt{i}, \texttt{j}, or \texttt{x} are penalized, because they frequently occur as local variables and are not useful retrieval anchors. Test-scope chunks are also penalized for non-test location queries, so that test files containing the same API calls do not outrank the actual implementation. Finally, the reranker can return diagnostics, including the extracted query tokens, exact filenames, exact symbols, matched API terms, entity preferences, and the full list of reranked candidates.

\subsection{LLMAgent}
\texttt{LLMAgent} does a conserative cleanup after retrieval in \texttt{\_filter\_final\_chunks()}. It does not rerank chunks again but focuses on filtering chunks which are too weak to show to the LLM, and asks if there is already better evidence from the same location.

A chunk gets a final-context quality score in \texttt{llm\_agent.py}. It gets positive points for:
\begin{enumerate}
    \item having a generated explanation
    \item having a leading comment
    \item having include paths or referenced files
    \item being high-level context: \texttt{file\_level}, \texttt{module\_level}, or \texttt{call\_chain\_level}
    \item having enough real code content
\end{enumerate}

Negative points are given to chunks if:
\begin{enumerate}
    \item \texttt{generated\_explanation\_status == "skipped\_low\_information"}
    \item being a trivial declaration, including empty or tiny declarations like \texttt{class X{};}
\end{enumerate}

Then the drop logic removes a chunk if all of the following are true:
\begin{enumerate}
    \item Its quality score is low, currently \texttt{<=0}
    \item It is either marked \texttt{skipped\_low\_information} or is a trivial declaration
    \item There is another chunk from the same local context
    \item That other chunk has a good enough score, at least 2.
    \item That other chunk beats it by at least 3 points.
\end{enumerate}

So the practical effect of the LLMAgent is: if retrieval returns both a useless \texttt{class FFT {};} declaration and a richer \texttt{FFT} file-level or method chunk, the agent may drop the empty declaration before building the prompt.

In general \texttt{LLMAgent.py} orchestrates the connection between the reranked retrieved context and the LLM, after filtering it calls up \texttt{build\_context()} to turn selected chunks into a text context string. This assures that from all context in the chunk a systematic and structural text is formed. This string gets inserted in the prompt and sent to the LLM.

\subsection{Prompting and Inference}
\label{sec:prompting-and-inference}

The project uses local LLM calls in several different stages of the pipeline. Therefore, it does not rely on a single prompt template, but defines separate prompt modes for different inference tasks. The first LLM calls already occur during retrieval preparation, where parsed entities are enriched with generated explanations. For symbol-level chunks, this is handled by the \texttt{EntityExplanationGenerator}; for higher-level entities, the \texttt{file\_level\_entity\_builder.py}, \texttt{module\_level\_entity\_builder.py}, and \texttt{call\_chain\_entity\_builder.py} modules use dedicated prompt templates together with \texttt{LLMWrapper} to generate retrieval-oriented entity explanations.

The prompt templates are defined in \texttt{prompt\_templates.py}. Each prompt mode is designed for a different type of context: local code entities, file summaries, module summaries, call-chain neighborhoods, or the final RAG answer. The active prompt modes are configured in \texttt{runtime\_settings.json}, while the complete prompt templates are listed in Appendix~\ref{app:prompt-templates}.

\begin{table}[H]
\centering
\begin{tabularx}{\linewidth}{>{\ttfamily}l X X}
\hline
\textbf{Prompt mode} & \textbf{Entity / use case} & \textbf{Prompting technique} \\
\hline
general & One parsed code entity or documentation section & Local entity explanation with a fixed structured schema \\
file\_level & Whole file with parsed symbols & Retrieval-friendly synthesis from structured file facts \\
file\_level\_fallback & Whole file without parsed symbols & Conservative fallback based on raw file evidence \\
module\_level & Folder or module summary & Aggregation of file-level summaries and module responsibilities \\
call\_chain & Callable plus incoming and outgoing calls & Local call-graph neighborhood explanation \\
retrieval\_answer & Final RAG answer, older style & Broad answer synthesis with several reasoning sections \\
retrieval\_answer\_v2 & Final RAG answer, evaluation style & Strict grounded question answering with constrained output \\
\hline
\end{tabularx}
\caption{Prompt modes and their associated prompting techniques.}
\label{tab:prompt-modes-techniques}
\end{table}

An important implementation detail is that prompt templates are part of the runtime manifest. The system computes prompt signatures for the active templates, so changing a prompt can invalidate previously generated artifacts and trigger a rebuild. This ensures that stored explanations remain consistent with the prompt version that produced them.

During final inference, \texttt{LLMAgent} formats the retrieved chunks into a structured context before calling the answering model. Each chunk is represented with metadata such as entity level, symbol name, path, namespace, generated explanation, references, and structural relationships. The final answer prompt therefore receives both semantic explanations and code-structure signals, while the prompt itself constrains the model to answer only from the retrieved context.

\section{Implementation and Runtime Configuration}
\label{sec:impl}
Since the architecture of this RAG-Scientific-Code-Agent separates retrieval preparation from the inference compeletely. It behaves fundamentally different than approaches in which retrieval preparation and inference are tightly coupled at query time. The main idea behind the architecture decision is to move as much computationally expensive work as possible into an offline pre-processing stage, performed once, when initializing the agent to a new codebase.

This is useful since codebase agents, are mainly used repeatedly within the same project environment, even though established models are designed to be adaptable to different codebases. In the proposed architecture, the system first ingests the target codebase and generates LLM-friendly retrieval context offline, using comparatively strong computational resources. The resulting vector store can then be reused during inference. As a result, the online answering stage can rely on smaller or more efficient models, because much of the code understanding has already been encoded into generated explanations, structural metadata, and retrieval-ready entities.

This section discusses the advantages and disadvantages of this architectural decision and documents the intended way of using, running and adapting the program to the ippl or other codebase environments.

\subsection{Configuration and Runtime Setup}

The configurable architecture of the software allows several important parameters for ingestion, embedding, prompting, retrieval, and model selection to be adjusted without changing the source code directly. These parameters are collected in \texttt{config/runtime\_settings.json}. This file therefore acts as the central runtime configuration of the pipeline and defines how the codebase is parsed, which models are used, how explanations are generated, and how retrieved chunks are ranked before they are passed to the final answering model.

For embedding, the active backend is \texttt{ollama}, using
\texttt{nomic-embed-text}~\cite{nussbaum2024nomic} as the embedding model. The
embedder also supports a sentence-transformer backend through the
\texttt{sentence\_transformer} option, with \texttt{BAAI/bge-code-v1} listed as
the configured sentence-transformer model. In the current evaluation setup,
however, embeddings are generated locally through Ollama.

\begin{lstlisting}
  "embedding": {
    "backend": "ollama",
    "ollama_model": "nomic-embed-text",
    "sentence_transformer_model": "BAAI/bge-code-v1"
  },
\end{lstlisting}

The configuration also defines the model roles used throughout the pipeline. The chunk-level explanation model is set to \texttt{qwen2.5-coder:32b}, while the file-level, module-level, call-chain, and final answer models use \texttt{qwen2.5-coder:32b-instruct-q4\_K\_M}. This separates the model used for symbol-level code explanations from the instruction-tuned model used for higher-level summaries and final question answering.

\begin{lstlisting}
    "models": {
    "answer_model": "qwen2.5-coder:32b-instruct-q4_K_M",
    "chunk_explanation_model": "qwen2.5-coder:32b",
    "file_level_model": "qwen2.5-coder:32b-instruct-q4_K_M",
    "module_level_model": "qwen2.5-coder:32b-instruct-q4_K_M",
    "call_chain_model": "qwen2.5-coder:32b-instruct-q4_K_M"
  },
\end{lstlisting}

Prompt behavior is also configured in \texttt{runtime\_settings.json}. The final answering stage uses \texttt{retrieval\_answer\_v2}, while the preprocessing stages use dedicated prompt modes for chunk explanations, file-level summaries, module-level summaries, and call-chain summaries. This makes it possible to change the prompting strategy for one stage without affecting the others.

\begin{lstlisting}
    "prompts": {
    "answer_prompt_mode": "retrieval_answer_v2",
    "chunk_explanation_prompt_mode": "general",
    "file_level_prompt_mode": "file_level",
    "file_level_fallback_prompt_mode": "file_level_fallback",
    "module_level_prompt_mode": "module_level",
    "call_chain_prompt_mode": "call_chain"
  },
\end{lstlisting}

The retrieval section of the configuration controls how aggressively candidates are selected and reranked. Parameters such as \texttt{candidate\_k}, \texttt{supplementary\_k}, \texttt{primary\_same\_symbol\_cap}, and \texttt{primary\_same\_file\_cap} influence how many chunks are considered and how much duplicate context from the same file or symbol is allowed. The weights \texttt{lexical\_metadata\_weight} and \texttt{entity\_target\_weight} control the importance of metadata overlap and query-intent preferences inside the reranking score.

\begin{lstlisting}
    "retrieval": {
    "candidate_k": 20,
    "supplementary_k": 3,
    "supplementary_candidate_k": 10,
    "primary_score_relative_floor": 0.4,
    "primary_score_gap_threshold": 8.0,
    "primary_score_absolute_floor": 1.25,
    "primary_same_symbol_cap": 1,
    "primary_same_file_cap": 2,
    "dominant_file_ratio": 0.75,
    "lexical_metadata_weight": 0.4,
    "entity_target_weight": 0.7,
    "debug_enabled_by_default": false
  }
\end{lstlisting}

To adapt the system to a different codebase, the most important setting is \texttt{raw\_data\_path}, which should point to the new source-code directory. Depending on the target project, the ingestion settings may also need to be adjusted, for example by changing the preferred parser, the maximum chunk size, or the entity types for which explanations should be generated. The model settings can be changed if different local models are available, and the retrieval parameters can be tuned if the new codebase requires broader or narrower context selection. After such changes, the vector store should be rebuilt with a clean rebuild so that the stored embeddings, generated explanations, and manifest match the new configuration.

\subsection{Execution on the Cluster}

When initializing the project for IPPL or for another codebase, the vector store has to be built from scratch. This initial build is the most computationally expensive stage of the pipeline, because the system parses the codebase, generates LLM-based entity explanations, computes embeddings, and stores the resulting chunks in the vector store. For this reason, the preprocessing step should be executed on a compute node with sufficient GPU resources.

In this project, the preprocessing job is submitted through SLURM using the command:

\begin{lstlisting}
FORCE_CLEAN_REBUILD=1 sbatch job.sh
\end{lstlisting}

The flag \texttt{FORCE\_CLEAN\_REBUILD=1} ensures that previously generated artifacts are removed and that the vector store is rebuilt from scratch. This is useful when changing important configuration parameters, such as the selected models, prompt templates, parsing strategy, retrieval settings, or the underlying source code that should be indexed.

The vector store used in this project was generated on the \texttt{gmerlin6} cluster, using the \texttt{gwendolen} partition. The submitted job requests one GPU and was executed on nodes equipped with NVIDIA A100 GPUs with 40 GB of GPU memory. These resources are especially important during the explanation-generation stage, where local LLM inference is repeatedly applied to parsed code entities. Once the vector store has been generated, the expensive preprocessing artifacts can be reused, and later inference runs no longer require rebuilding the full index unless the configuration or input codebase changes.

\subsection{Architectural Advantages and Limitations}

The architecture of the RAG-Scientific-Code-Agent differs from general-purpose coding agents such as Codex~\cite{openai2025codex} or Claude Code~\cite{anthropic2025claudecode}. Codex and Claude Code are designed as interactive agentic coding systems: they can inspect a repository, reason over the current task, edit files, run commands, execute tests, and iterate based on the results. In contrast, the system developed in this project focuses on retrieval-augmented code understanding. Its main architectural decision is to separate expensive retrieval preparation from later inference. The codebase is parsed, enriched with generated explanations, embedded, and stored in a vector store before the user asks questions.

This design has several advantages. First, expensive work is moved into an offline preprocessing stage. LLM-based chunk, file, module, and call-chain explanations only have to be generated once for a given codebase version and configuration. During inference, the system can reuse this prepared retrieval context instead of repeatedly asking the model to rediscover the structure of the repository. Second, the generated vector store is specialized for the target codebase. This is especially useful for scientific C++ projects, where relevant information is often distributed across templates, headers, implementation files, namespaces, and build or test directories. Third, the architecture improves reproducibility: the runtime settings, prompt modes, model names, and generated artifacts can be stored and compared across evaluation runs. Finally, because much of the code understanding is already encoded into retrieval-ready explanations and metadata, the final answering step can be performed with a smaller or more efficient local model than would otherwise be required.

However, this architecture also has disadvantages. The initial preprocessing step is computationally expensive and can require a powerful GPU node, especially when explanations are generated for many code entities. The vector store can also become stale: if the source code, parser, prompt templates, or model configuration changes, the index may need to be rebuilt. In addition, the system depends strongly on the quality of parsing, chunking, embedding, reranking, and structural expansion. If an entity is not extracted correctly or not retrieved, the final LLM cannot reason about it. Compared with agentic systems such as Codex~\cite{openai2025codex} or Claude Code~\cite{anthropic2025claudecode}, the system is also less autonomous. It does not directly modify code, run tests, inspect errors, or iteratively verify its own answer by executing commands.

The comparison therefore shows a trade-off between specialization and autonomy. Codex and Claude Code are more flexible for active software engineering tasks such as debugging, refactoring, writing tests, and creating pull requests. Their strength is the interactive agentic loop: gathering context, taking actions, observing results, and adapting. The RAG-Scientific-Code-Agent, on the other hand, is optimized for repeated question answering over a relatively stable scientific codebase. Its strength is controlled, reproducible, and domain-specific retrieval. It is therefore better suited for explaining and navigating a prepared codebase, while Codex- or Claude-style agents are better suited for changing and verifying code during active development.
\section{Evaluation}
\label{sec:eval}

\subsection{Experimental Setup}
We evaluate answer accuracy, faithfulness, and development utility using 100
benchmark questions generated with GPT-5.4 (Codex)~\cite{openai2026gpt54} after providing the model
with the complete IPPL codebase. The questions span the eleven categories in
\Cref{tab:evaluation-question-categories}; the complete set is provided in
\Cref{app:evaluation-questions}.

GPT-5.5~\cite{openai2026gpt55} is used as an LLM judge to compare each generated answer with its
reference answer. The judge assigns 1 point to a correct answer, 0.5 to a
partially correct answer, and 0 to an incorrect answer, and records a rationale
for answers that are not fully correct. LLM-based judging provides a scalable
evaluation mechanism but is known to exhibit systematic biases~\cite{zheng2023llmjudge}. This automatic evaluation design is a
potential source of bias, as discussed in \Cref{sec:discussion}.

\begin{table}[H]
\centering
\begin{tabularx}{\linewidth}{l r X}
\hline
\textbf{Question category} & \textbf{Count} & \textbf{What the questions ask} \\
\hline
\texttt{algorithm} & 12 & Algorithmic behavior of core numerical methods, such as FFT Poisson solvers, CG iteration, interpolation, FFTs, and particle-field operations. \\
\texttt{api\_usage} & 12 & How to use IPPL APIs in code, for example creating fields, meshes, boundary conditions, particle classes, initialization/finalization, and solver setup. \\
\texttt{boundary\_and\_halo} & 5 & Boundary condition types, differences between boundary conditions, ghost-cell population, halo exchange, and absorbing boundaries. \\
\texttt{build\_and\_install} & 5 & Build requirements, third-party dependencies, CMake/GPU configuration, Kokkos setup, and cluster installation instructions. \\
\texttt{class\_responsibility} & 13 & Responsibilities of important classes such as \texttt{BareField}, \texttt{Field}, \texttt{FieldLayout}, \texttt{ParticleBase}, solver classes, and layout classes. \\
\texttt{data\_flow} & 9 & How data moves through the codebase, including particle-to-grid charge deposition, grid-to-particle field gathering, halo exchange, particle migration, and solver data flow. \\
\texttt{definition\_location} & 15 & Where specific implementations live in the source tree, such as MPI setup, Kokkos initialization, FFT initialization, load balancing, and solver implementations. \\
\texttt{file\_purpose} & 13 & The role of important source/header files, for example \texttt{Ippl.h}, \texttt{Ippl.cpp}, \texttt{FFT.hpp}, \texttt{BareField.h}, and mini-app files. \\
\texttt{numerical\_meaning} & 3 & Physical or numerical interpretation of methods, such as what Poisson solvers compute, what FDTD fields mean, and tradeoffs between FFT and CG solvers. \\
\texttt{parallelism\_and\_kokkos} & 9 & How IPPL uses Kokkos, MPI, execution spaces, \texttt{Kokkos::View}, parallel loops, mixed precision, and CPU/GPU portability. \\
\texttt{testing\_and\_workflow} & 4 & Development workflow, naming conventions, profiling, math-function conventions, and SLURM job templates. \\
\hline
\textbf{Total} & \textbf{100} & Complete current evaluation benchmark. \\
\hline
\end{tabularx}
\caption{Categories in the current evaluation benchmark and the type of questions each category contains.}
\label{tab:evaluation-question-categories}
\end{table}

For evaluating different models the pipeline uses one fixed model setup for the "explanation generation" and chunk embeddings.

\begin{table}[H]
\centering
\begin{tabular}{ll}
\hline
\textbf{Pipeline role} & \textbf{Model used in latest SSH evaluation} \\
\hline
Embedding & \texttt{nomic-embed-text} \\
Chunk explanation model & \texttt{qwen2.5-coder:14b} \\
File-level model & \texttt{qwen2.5-coder:32b-instruct-q4\_K\_M} \\
Module-level model & \texttt{qwen2.5-coder:32b-instruct-q4\_K\_M} \\
Call-chain model & \texttt{qwen2.5-coder:32b-instruct-q4\_K\_M} \\
\hline
\end{tabular}
\caption{Models used in the latest SSH evaluation run on the Gwendolen partition.}
\label{tab:ssh-evaluation-models}
\end{table}

The only varying component is the answering language model, which receives the
retrieved context, code, metadata, and previous LLM generations. The evaluation
includes models from the Qwen2.5~\cite{qwen2024qwen25},
Qwen3.5~\cite{qwen2026qwen35}, and Gemma 4~\cite{gemma2026gemma4} families. The
tested configurations are listed below.

\begin{table}[H]
    \centering
    \begin{tabular}{c|ccc}
        \textbf{Model} & \textbf{Parameters} & \textbf{Context Window} & \textbf{Memory Sizes}\\
        \hline
        Qwen2.5-coder-4\_K\_M & 7B  & 32K  & $\sim$7\,GB\\
        Qwen2.5-coder-4\_K\_M & 14B & 32K  & $\sim$12\,GB\\
        Qwen2.5-coder-4\_K\_M & 32B & 32K  & $\sim$24\,GB\\
        Qwen3.5               & 32B & 256K & $\sim$28\,GB\\
        Qwen3.5               & 9B  & 256K & $\sim$9\,GB\\
        Gemma4                & 31B & 256K & $\sim$24\,GB\\
        Gemma4                & 12B & 256K & $\sim$11\,GB\\
    \end{tabular}
    \caption{Retrieval Answering models on which we evaluate the 100 Benchmark questions. Memory sizes are approximate VRAM requirements for the individual model. }
    \label{tab:answering-models}
\end{table}

In order to test retrieval quality directly the benchmark questions are also tested against a different architecture, for this the evaluation ran over the claude code ~\cite{anthropic2025claudecode} retrieval architecture with some of the same models used before in the test run with the RAG architecture.

\begin{table}[H]
    \centering
    \begin{tabular}{c|ccc}
        \textbf{Model} & \textbf{Parameters} & \textbf{Context Window} & \textbf{Memory Sizes}\\
        \hline
        Claude Code (Qwen2.5-coder-4\_K\_M) & 7B & 32K & $\sim$7\,GB\\
        Claude Code (Qwen2.5-coder-4\_K\_M)  & 32B & 32K & $\sim$24\,GB\\
        Claude Code (Gemma4) & 12B  & 256K & $\sim$11\,GB
    \end{tabular}
    \caption{Models used in comparison for the Claude Code (Anthropic)~\cite{anthropic2025claudecode} architecture on the benchmark questions for comparison to our architecture. The pipeline here uses the Claude Code architecture to find the answer inside the ippl codebase only using the specified smaller model. }
\end{table}

\subsection{Results}
\label{sec:results}
Evaluating the different models based on the explained setup in the previous section, showed interesting results. The older, well-established Qwen2.5 models underperformed with a range from 0.665 to 0.720. The Gemma4 models were also ranked among these values; however, the Qwen3.5 model is the clear winner out of the three. What was expected and can be clearly seen in the resulting data is that newer models or higher parameter counts tend to increase latency. Note that all latency was measured on the merlin6 ssh.

\begin{table}[H]
  \centering
  \makebox[\textwidth][c]{%
  \begin{tabular}{cc|cccc|c}
    \toprule
    Model & Parameters & Correct [1] & Partial [0.5] & Incorrect [0] & Mean Latency & Average Score \\
    \midrule
    Qwen2.5 & 7B   & 59 & 26 & 15 & 3.54s  & $0.720 \pm 0.037$ \\
    Qwen2.5 & 14B  & 60 & 17 & 23 & 6.95s  & $0.685 \pm 0.042$ \\
    Qwen2.5 & 32B  & 55 & 23 & 22 & 14.13s & $0.665 \pm 0.041$ \\
    Qwen3.5 & 9B   & 70 & 19 & 11 & 18.14s & $\mathbf{0.795 \pm 0.034}$ \\
    Qwen3.5 & 32B  & 65 & 23 & 12 & 30.91s & $0.765 \pm 0.035$ \\
    Gemma4  & 12B  & 56 & 23 & 21 & 38.72s & $0.675 \pm 0.040$ \\
    Gemma4  & 31B  & 63 & 16 & 21 & 41.00s & $0.710 \pm 0.041$ \\
    \bottomrule
  \end{tabular}
  }
  \caption{Main results for 7 different models, based on 100 benchmark questions. Values after $\pm$ denote the standard error of the mean over $n = 100$ questions. Evaluated on Merlin6 via SSH by running \texttt{job\_eval.sh}.}
  \label{tab:main-results}
\end{table}

\subsubsection{Claude Code Comparison Results}
The results of a frontier retrieval architecure such as Claude Code ~\cite{anthropic2025claudecode} resulted in a smaller average score than for the Rag-Scientific code agent implemented in this project. The two instruct models of Qwen2.5-q4\_K\_M showed a score of $0.520 \pm 0.038$ for the smaller 7B model and an average score of $0.560 \pm 0.036$ for the 32B model. The Agentic loop alongside the Gemma4 model scored higher with an average score of $0.665 \pm 0.040$. 

\begin{table}[H]
  \centering
  \begin{tabular}{l|ccc}
    \toprule
    & \multicolumn{3}{c}{Claude Code} \\
    \cmidrule(lr){2-4}
    Metric & Qwen2.5-q4\_K\_M & Qwen2.5-q4\_K\_M  & Gemma4 \\
    Parameters & 7B & 32B & 12B \\
    \midrule
    Correct       & 30 & 33 & 53 \\
    Partial       & 44 & 46 & 27 \\
    Incorrect     & 26 & 21 & 20 \\
    Mean Latency  & 6.2s  &  40.8s  &  92.4s  \\
    \hline
    Average Score &$ 0.520 \pm 0.038$ & $0.560 \pm 0.036$  & $\mathbf{0.665 \pm 0.040}$ \\
    \bottomrule
  \end{tabular}
  \caption{Comparison results using the Claude Code~\cite{anthropic2025claudecode} pipeline with Qwen2.5 and Gemma4~\cite{gemma2026gemma4} models for comparison with the RAG pipeline.}
  \label{tab:claude-code-results}
\end{table}

What is notable when looking at the results is that for both instruct models most answers are only partially correct, with 44/100 questions 'Partial' for the 7B model and 46/100 for the 32B model. Whereas the Gemma4 model stays around the same amount of 20/100 'Incorrect' answers as the other two, but only marks 27/100 answers as 'Partial'. 

\subsubsection{Category Breakdown}
Based on the seven evaluations we generated for our pipeline, the strongest category overall is API Usage. Across all models, it had 69 correct, 10 partial, and only 5 incorrect out of 84 total answers, giving an aggregate average score of 0.881. This means the models were usually good when the question asked how an API, class, method, or usage pattern works.

The aggregated breakdown across all evaluated models shows: 

\begin{table}[H]
\centering
\begin{tabular}{lrrrr|r}
\hline
\textbf{Category} & \textbf{Total answers} & \textbf{Correct} & \textbf{Partial} & \textbf{Incorrect} & \textbf{Avg. score}\\
\hline
API Usage & 84 & 69 & 10 & 5 & $0.881 \pm 0.030$\\
Definition Location & 105 & 81 & 2 & 22 & $0.781 \pm 0.040$\\
Class Responsibility & 91 & 63 & 13 & 15 & $0.764 \pm 0.039$\\
File Purpose & 91 & 54 & 28 & 9 & $0.747 \pm 0.035$\\
Boundary And Halo & 35 & 20 & 12 & 3 & $0.743 \pm 0.056$\\
Data Flow & 63 & 38 & 17 & 8 & $0.738 \pm 0.050$\\
Algorithm & 84 & 39 & 33 & 12 & $0.661 \pm 0.039$\\
Parallelism And Kokkos & 63 & 27 & 23 & 13 & $0.611 \pm 0.049$\\
Testing And Workflow & 28 & 16 & 2 & 10 & $0.607 \pm 0.090$\\
Numerical Meaning & 21 & 11 & 1 & 9 & $0.548 \pm 0.109$\\
Build And Install & 35 & 10 & 6 & 19 & $0.371 \pm 0.073$\\
\hline
\end{tabular}
\caption{Aggregated category-level performance across all evaluated models. Values after $\pm$ denote the standard error of the mean.}
\label{tab:aggregated-category-performance}
\end{table}

The categories that worked well were API Usage, Definition Location, Class Responsibility, File Purpose, Boundary And Halo, and Data Flow. These all scored around 0.74 or higher, meaning the models often found the right file, class, function, or conceptual role. For example, Definition Location had 81 correct out of 105, and File Purpose had only 9 incorrect out of 91, even though many answers were partial.

The weakest category was Build And Install, with only 10 correct out of 35 and 19 incorrect, for an average score of 0.371. That suggests the models struggled with setup, installation, CMake/build instructions, and workflow details. Numerical Meaning was also weak, with 11 correct, 1 partial, and 9 incorrect out of 21, average 0.548, although this category has fewer questions. Parallelism And Kokkos and Testing And Workflow were also mixed, both around 0.61, likely because those questions require more cross-file reasoning and exact understanding of project conventions rather than simply locating a symbol.

\begin{table}[H]
\centering
\begin{tabular}{lrrrr|r}
\hline
\textbf{Category} & \textbf{Questions} & \textbf{Correct} & \textbf{Partial} & \textbf{Incorrect} & \textbf{Avg. Score}\\
\hline
File Purpose & 13 & 9 & 4 & 0 & $0.846 \pm 0.067$\\
Definition Location & 15 & 13 & 0 & 2 & $0.867 \pm 0.091$\\
Class Responsibility & 13 & 9 & 2 & 2 & $0.769 \pm 0.108$\\
Algorithm & 12 & 8 & 3 & 1 & $0.792 \pm 0.096$\\
Data Flow & 9 & 5 & 3 & 1 & $0.722 \pm 0.121$\\
API Usage & 12 & 11 & 1 & 0 & $0.958 \pm 0.042$\\
Parallelism and Kokkos & 9 & 6 & 2 & 1 & $0.778 \pm 0.121$\\
Boundary and Halo & 5 & 4 & 1 & 0 & $0.900 \pm 0.100$\\
Numerical Meaning & 3 & 2 & 0 & 1 & $0.667 \pm 0.333$\\
Build and Install & 5 & 1 & 2 & 2 & $0.400 \pm 0.187$\\
Testing and Workflow & 4 & 2 & 1 & 1 & $0.625 \pm 0.240$\\
\hline
\end{tabular}
\caption{Category breakdown for the Qwen3.5 9B model. Values after $\pm$ denote the standard error of the mean.}
\label{tab:category-breakdown}
\end{table}

\subsection{Discussion}
\label{sec:discussion}
A key observation in \Cref{tab:main-results} is the decrease in accuracy when increasing model size. It is clearly visible for the Qwen2.5-coder model, where the 32B model performs with an average score of 0.055 lower than its smaller 7B model. Notable to see is, that this is no exception, the Qwen3.5 32B model is 0.03 lower than its 9B version. The only model which doesn't follow this rule is the gemma4 model in which an increase in Average Score of 0.035 can be seen when increasing the parameter size from 12B to 31B. 

The reason for this phenomenon can vary and is not clear. What is clear is that the benchmark rewards short, grounded, repository-specific answers. Large models often give broader answers. If the model includes irrelevant details, misses the exact source location, or wanders from the question, the score drops. 

What can be accounted for scoring accuracy are the general generation settings, same prompt, temperature, context length, and max tokens do not affect every model equally. A setting that works well for Qwen3.5 9B may be suboptimal for Gemma4 31B or Qwen2.5 32B model. 

Generally the results in \Cref{tab:main-results} show that model family matters more than size, a 9B model from a newer or better-aligned family can beat a 31B model from another family. In the results Qwen3.5 9B scored higher than Gema4 31B, likely because it was better suited to code/RAG-style question answering. 

This is also again underlined when comparing Qwen3.5 9B with the 32B model which only differences 0.03 in the benefit of the 9B model. When comparing the 9B model to other models of similar size, the pattern
holds: within the smaller size class, the newer Qwen3.5 family clearly
outperforms the older Qwen2.5-coder generation, confirming that architectural
and alignment improvements outweigh raw scale for this task.

A second, more important comparison concerns the retrieval architecture
itself. \Cref{tab:claude-code-results} shows that the same
models perform substantially worse when embedded in the Claude Code retrieval
pipeline than in our dedicated RAG pipeline. The Qwen2.5 7B model drops from
an average score of 0.720 in our pipeline to 0.520 under Claude Code.
This is a striking result: the agentic retrieval loop, which is designed
around very large frontier models, does not transfer well to small local
models. Without a strong model to drive iterative file inspection and tool
use, the agentic approach appears to retrieve noisier or less targeted
context, whereas our offline-prepared retrieval context is already condensed
into grounded, repository-specific explanations that a small model can use
directly.

This supports the central hypothesis of the project: when the answering model
is small, the quality of the prepared retrieval context matters more than the
sophistication of the retrieval-time agent loop. Front-loading code
understanding into the ingestion stage effectively compensates for the
limited reasoning capacity of small models.

Two caveats should be noted when interpreting these results. First, the
benchmark answers and the calibration baseline were generated by OpenAI
models (GPT-5.4 for question generation, GPT-5.5 for scoring). An LLM judge
may carry subtle stylistic preferences---for example, favouring concise,
structured answers---which could systematically advantage or disadvantage
certain answering models independently of factual correctness. Second, the
benchmark consists of 100 questions over a single codebase (IPPL), so the
absolute scores should be read as indicative rather than definitive. A larger
multi-codebase benchmark and, ideally, partial human validation of the LLM
judgments would strengthen these conclusions.

\section{Conclusion}
\label{sec:conclusion}
This study investigated whether a useful coding agent can be built
around small, locally executed open-source models. The implemented
RAG scientific-code agent demonstrates that this is feasible: by separating an
expensive offline ingestion stage from a lightweight online answering stage,
the system enables small models to produce grounded, repository-specific
answers on the IPPL codebase. The strongest configuration, a 9B model,
achieved an average score of $0.795 \pm 0.034$ on a 100-question benchmark, outperforming
larger models within the same pipeline as well as the same models embedded in
an agentic Claude Code retrieval architecture.

Two findings stand out. First, for this task model family and retrieval
quality matter more than parameter count---newer, better-aligned model
families consistently outperformed larger but older ones. Second, the
offline-prepared retrieval context appears to be precisely what allows small
models to remain competitive, since it condenses code understanding into
explanations and structural metadata that a small model can use without
extensive reasoning of its own.

It is worth emphasizing what the system is and is not. The current architecture
is based solely on semantically explaining a codebase rather than acting as an
all-round code agent; it does not edit files, run tests, or iteratively verify
its own output. The result is therefore best understood as a privacy-preserving
``code teacher'' that runs locally and free of charge, allowing anyone to
explore how a scientific codebase works while keeping their data entirely on
their own infrastructure. For active software engineering tasks, additional
tooling would be required.

The central trade-off of the design is its greatest strength and its greatest
weakness. Front-loading code understanding into the ingestion stage reduces
inference cost and lets small models perform well, but any change to the
codebase requires re-running the costly ingestion pipeline. This makes the
agent best suited to relatively stable codebases that are queried repeatedly,
such as those found in educational settings---universities, schools, or
corporate onboarding---where new students or employees can rely on a safe,
local, and free tool to learn how a large scientific codebase is structured
and how it works.

\subsection{Future Work}
Although the project is already substantial, several parts of the pipeline remain to be improved or tested. Any such improvement should be evaluated carefully and in context, since changes made without proper evaluation could just as easily degrade the pipeline rather than improve it.

Starting at the beginning of the pipeline, the embedding stage, one improvement would be clearer formatting and symbolisation of code blocks, for example marking them explicitly with backticks:
\begin{lstlisting}
-- CHUNK START --
    Document Section: User Authentication Handler
    Semantic Context: This function validates the incoming JSON payload.
    Code:
```python
    def verify_age(payload):
        age = payload.get("age", 0)
        return isinstance(age, int) and age >= 18
```
-- CHUNK END --
\end{lstlisting}
This, along with other experiments in chunk formatting, could improve both retrieval and context quality, since chunk formatting plays an important role in how well an LLM understands the input.

A further set of improvements involves swapping individual software components in the pipeline, such as the embedder, the retriever, or the vector store, as well as adding a pre-built re-ranking library. For re-ranking, \texttt{BAAI/bge-reranker-v2-m3} or \texttt{BAAI/bge-reranker-large} would be candidates for evaluation. Cross-encoder rerankers add computation to retrieval, but they may recover text chunks that describe a code concept accurately despite receiving a low initial vector-similarity score.

Finally, the agent could be extended with external tooling such as \texttt{opencode} to act as an implementation helper rather than purely a learning tool. This would likely require a revision of the codebase first, so it should be treated as a longer-term goal. In the nearer term, the higher priority is constant evaluation while experimenting with alternative frameworks, retrievers, models, and embedders.

\section*{Data and Code Availability}
The implementation, evaluation material, and instructions needed to reproduce
the reported experiments are available at
\url{https://github.com/nobileaaron/rag-scientific-code-agent}.

\section*{Acknowledgements}
The authors thank the contributors to the IPPL codebase and the computing
staff who supported the evaluation runs.

\bibliographystyle{unsrt}
\bibliography{references}

\appendix
\section{Evaluation Protocol}
\label{app:extra}
\begin{lstlisting}[
  keywordstyle=\color{black},
  stringstyle=\color{black},
  commentstyle=\color{black},
  language={}
]
    You are evaluating the currently selected Claude Code model on the IPPL benchmark.

Task:
Read `docs/evaluations/eval_questions_v2.json`, answer every question using only evidence from this repository, especially `data/raw/ippl` and repository documentation, and save the results as a JSON file under `docs/evaluations/answers/`.

Important requirements:
- Do not modify source code or configuration files.
- Create `docs/evaluations/answers/` if it does not exist.
- Use the environment variable `CLAUDE_EVAL_MODEL_LABEL` as the run label. If it is not set, use `unknown_model`.
- Save to `docs/evaluations/answers/claude_code_<run_label>_<UTC timestamp>.json`.
- Write progress incrementally after each answered question with `"run_complete": false`, so partial progress is preserved if the run stops.
- At the end, rewrite the same file with `"run_complete": true`.
- Use this output schema:

{
  "schema": "claude-code-direct-eval/v1",
  "run_complete": false,
  "run_label": "...",
  "run_started_at_utc": "...",
  "run_finished_at_utc": null,
  "questions_source": "docs/evaluations/eval_questions_v2.json",
  "answering_instructions": "...",
  "answers": [
    {
      "id": "...",
      "category": "...",
      "question": "...",
      "answer": "...",
      "evidence_files": ["..."],
      "error": null,
      "latency_seconds": 0.0
    }
  ]
}

Answering rules:
- For each question, give a concise technical answer.
- Ground the answer in concrete files, classes, functions, or documentation sections when possible.
- Include relevant evidence file paths in `evidence_files`.
- If the repository evidence is insufficient, say so explicitly in the answer instead of guessing.
- Do not use external web search.
- Do not include long copied code blocks unless necessary.
- Keep answers comparable across runs: use the same style and level of detail for every model.

Implementation:
Use shell/Python as needed to parse the JSON and write the output file, but generate the actual answer content yourself from repository inspection. Proceed through all questions without asking me for confirmation.
\end{lstlisting}

\section{Prompt Templates}
\label{app:prompt-templates}
The complete system and user prompts for chunk explanation, file- and
module-level summarization, call-chain summarization, and retrieval-based
answering are distributed with the reproducibility artifacts in
\texttt{appendices/prompt\_templates\_overview.txt} and in the public code
repository. They are omitted here because several templates are long machine
inputs whose verbatim inclusion does not aid interpretation of the method.

\section{Evaluation Questions}
\label{app:evaluation-questions}
The complete benchmark is distributed as
\texttt{appendices/eval\_questions\_v2.txt} and with the public repository.
It contains 100 questions in the eleven categories summarized in
\Cref{tab:evaluation-question-categories}. Providing the artifact separately
keeps the manuscript concise while preserving the exact benchmark wording.

\end{document}